\documentclass[aps,prb,twocolumn,superscriptaddress,showpacs]{revtex4}

\usepackage[utf8]{inputenc}
\usepackage[english]{babel}
\usepackage{graphicx}
\usepackage{amsmath}
\usepackage{amssymb}
\usepackage{color}
\usepackage[abs]{overpic}

\usepackage{cancel}

\newcommand{\eS}{\mathcal{S}}

\newcommand{\eM}{\mathcal{M}}
\newcommand{\eN}{\mathcal{N}}
\newcommand{\eL}{\mathcal{L}}

\newcommand{\trace}{\mbox{Tr}}

\newcommand{\ie}{i.e.~}
\newcommand{\eg}{e.g.~}

\definecolor{darkgreen}{rgb}{0,0.5,0}

\begin{document}

\title{Ab-initio characterization of the quantum linear-zigzag transition using DMRG}

\author{Pietro Silvi}
\affiliation{Institut f\"ur Quanteninformationsverarbeitung,
  		Universit\"at Ulm, D-89069 Ulm, Germany}
\author{Tommaso Calarco}
\affiliation{Institut f\"ur Quanteninformationsverarbeitung,
  		Universit\"at Ulm, D-89069 Ulm, Germany}
\author{Giovanna Morigi}
\affiliation {Theoretische Physik, Universit\"at des Saarlandes,
		D-66123 Saarbr\"ucken, Germany}
\author{Simone Montangero}
\affiliation{Institut f\"ur Quanteninformationsverarbeitung,
  		Universit\"at Ulm, D-89069 Ulm, Germany}

\date{\today}

\begin{abstract}

Ions of the same charge inside confining potentials can form crystalline structures which can be controlled by means of
the ions density and of the external trap parameters. In particular, a linear chain of trapped ions exhibits a transition
to a zigzag equilibrium configuration, which is controlled by the strength of the transverse confinement.
Studying this phase transition in the quantum regime is a challenging problem, even when employing numerical methods to
simulate microscopically quantum many-body systems. Here we present a {\bf compact} analytical treatment
to map the original long-range problem into a short-range quantum field theory on a lattice.
We provide a complete numerical architecture,
based on Density Matrix Renormalization Group, to address the effective quantum $\phi^4$ model.
This technique is instrumental in giving a complete characterization of the phase diagram, as well as pinpoint the
universality class of the criticality.

\end{abstract}

\pacs{
61.50.-f, 
64.70.Tg, 05.30.Rt, 
05.10.-a. 
}

\maketitle

\section{Introduction}\label{sec:intro}

Wigner crystals~\cite{Wigcrys} composed of trapped and mutually-repelling ions are an outstanding prototype of tailored condensed matter \cite{Dubin_RMP,zig1,Bollinger:1,Drewsen:1,Bloch:2}. The high degree of control they offer makes them an ideal platform for quantum information
processing \cite{CiracZoller,WinelandQC,BlattQC}, quantum communication devices~\cite{Ioncommunicator,Yelin,Drewsen_EIT}
and quantum simulators~\cite{Feynman, Bloch:1,Schneider, Friedenauer,Monroe,Ionsimulator,PorrasCirac,Ultralewen}.
Moreover, they constitute a perfect playground for studying general, distinctive features of condensed-phase systems; above all, phase transitions and critical phenomena~\cite{Mussardo, Sachdev,zig1,Monroe,Bollinger:2,Drewsen:2,Mainz,Morigi:G.5,Shimshoni:2011a,Bermudez}.

\begin{figure}
 \begin{center}
 \begin{overpic}[width = \columnwidth, unit=1pt]{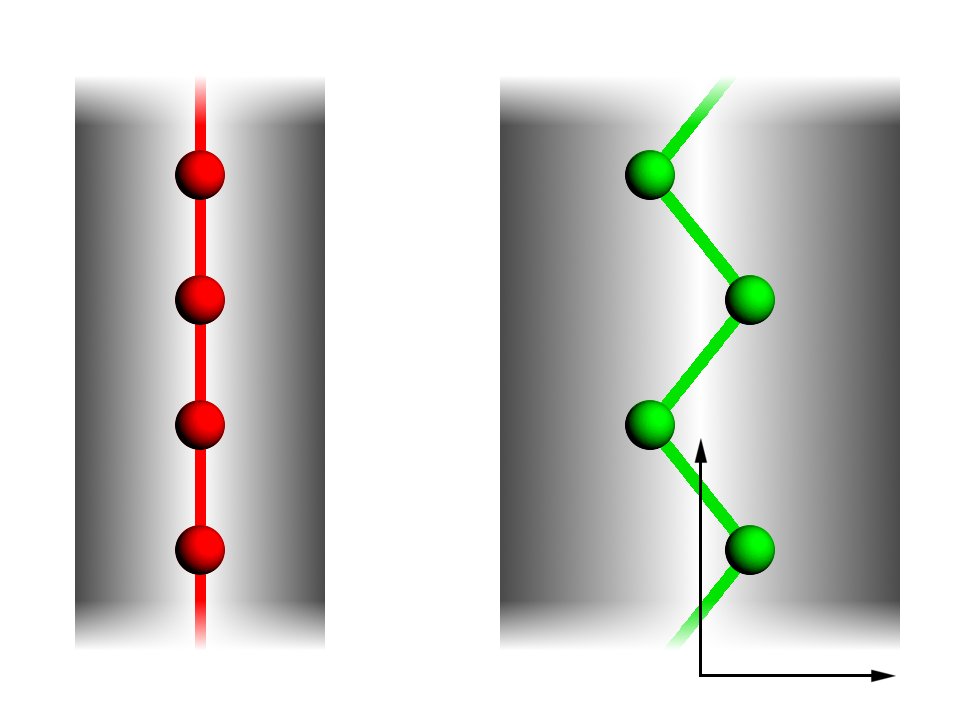}
  \put(200, 3){$y$}
  \put(167, 40){$x$}
 \end{overpic}
 \end{center}
\caption{ \label{fig:design} (color online)
Linear-zigzag instability in a chain of interacting atoms: (left) linear and (right) zigzag configurations. The transition is either controlled by changing the longitudinal interparticle spacing $a$ or the frequency $\omega_T$ of the transverse harmonic confinement~\cite{zig1,Morigi:G.5}.
The displacement of a particle from the axis $y = 0$ plays the role of local order parameter, discriminating between
the disordered (linear) phase and the ordered (zigzag) phase.
}
\end{figure}

One prominent example is the linear ion chain \cite{Dubin_RMP,zig1,Paul1}, which results from the interplay between long-range Coulomb repulsion and a highly anisotropic confinement due to an ion trap \cite{Ghosh}. This quasi-ordered structure can become unstable depending on the trap aspect ratio or on the ion density. Fig.~\ref{fig:design} illustrates the two equilibrium configurations: Here, zigzag order (right panel) becomes energetically favourable at lower transverse confinements \cite{zig1,Paul1,Dubin1993,Schieffer1993,Morigi:G.5}. While it was often argued in the literature that this structural instability is a continuous phase transition \cite{Dubin1993,Schieffer1993}, in Ref. \cite{Morigi:G.5} it was first demonstrated that it can be rigorously mapped to a Landau second-order phase transition in the appropriately defined thermodynamic limit
\cite{Morigi:G.7,DengPorrasCirac}. This structural instability gathered a lot of interest in the latest years as a laboratory system for studying quenches across critical points \cite{DeChiara,Baltrusch,Exp:KZ,Exp:KZ:2}, and also because it is believed to be a promising channel for transport and storage of quantum entanglement \cite{Ioncommunicator,Yelin}.

Quantum effects at the critical point have been theoretically studied in Ref. \cite{Retzker:2008} for small ion chains. In Refs. \cite{Meyer,Shimshoni:2011a} it has been argued that in the thermodynamic limit the linear-zigzag structural instability is a quantum phase transition, which in two ($1+1$) dimensions can be mapped to an Ising model in the transverse field, describing a ferromagnetic transition at zero temperature. This mapping was first proposed for Wigner crystals of electrons in quantum wires \cite{Meyer}, and then derived in Ref. \cite{Shimshoni:2011a,Shimshoni:2011b} using the emergent $\mathbb{Z}_2$ symmetry. In Ref. \cite{Shimshoni:2011b}, in particular, parameter regimes were estimated for which the quantum phase transition could be experimentally measured. It was noticed that, while for ions achieving the quantum critical region can be experimentally challenging, it could be more easily accessed with other kinds of strongly-correlated systems, for instance, dipolar atomic gases\cite{
Menotti} in elongated traps. In this context, we mention that theoretical studies of the linear-zigzag instability with ultracold dipolar systems appeared in Refs. \cite{Morigi:Dipoli,Altman}. Related phenomena were identified in arrays of dipolar tubes \cite{Kollath}. 

This framework motivates an accurate characterization of the quantum behavior at criticality, which shall provide ultimate evidence of the universality class of the quantum linear-zigzag instability and allow one to determine the parameter ranges where it can be experimentally measured. Starting from Ref. \cite{Morigi:G.5}, where the long-wavelength behavior at the instability was mapped to a $\phi^4$ model, the question can be posed in more general terms, namely, whether a Landau-Ginzburg model in $1+1$ dimensions belongs to the same universality class of the Ising model in a transverse field \cite{Cardy}. This problem was already numerically approached in Refs. \cite{Barma, MontecarloP4, Sugihara}, but only partial conclusions could be reached.

In this manuscript we address the quantum scenario of the linear-zigzag phase transition by means of Density Matrix Renormalization Group (DMRG)~\cite{White92}, a numerical technique tailored on correlated quantum many-body systems on a one-dimensional lattice~\cite{PrimoMPS}.
With this technique it is possible to address the quantum phase transition problem,
and verify that the linear-zigzag instability belongs to the universality class of the Ising model in a transverse field by determining the relevant critical exponents. Similarly, the simulation can quantify the quantum corrections to the classical linear-zigzag transition.

Here we take a special care in developing in full detail the theoretical framework as well as the numerical architecture employed.
The contribution given by this paper is threefold: first, we
provide for the first time a compact formulation for the mapping from a long-range
into a short-range linear-zigzag model at any order of expansion in the displacement.
Secondly, we give a robust and scientifically sound background to the results some of us previously presented
in Ref.~\cite{NostroANDP}, as well as expand that work by adding previously unreported comments.
Finally, we take a special care in reporting all the numerical expedients we adopt so that all of our
results will be fully reproducible.

The paper is organized as follows. In section \ref{sec:modmod} we introduce the model and we describe a quantitatively robust mapping of the system Hamiltonian into a simpler one, which can be easily tackled numerically. In section \ref{sec:simsim} we review the numerical strategies that we employ to tailor the effective
model into a DMRG architecture, as well as some techniques for quantum state analysis and data-processing. The phase diagram of the quantum linear-zigzag phase transition is sketched in section \ref{sec:phasedia}. In section \ref{sec:uniclass} we determine the relevant critical exponents. The conclusions are drawn in section \ref{sec:conclu}, while the appendix provides details of the mapping presented in section \ref{sec:modmod}.

\section{Quasi-1D Wigner crystal}\label{sec:modmod}

We consider a two-dimensional system of $L$ interacting atoms, trapped by a harmonic potential with frequency $\omega$ along the $y$ direction. The atoms are identical and have mass $M$. Typical distances between atomic ions are of mesoscopic scale \cite{zig1}, so that they are ultimately distinguishable. This allows us to write a first-quantization Hamiltonian despite the many-body character of the dynamics, which reads
\begin{multline} \label{eq:original}
 H = \sum_{j = 1}^{L}
 \left[ \frac{p_{x,j}^2 + p_{y,j}^2}{2M} + \frac{M \omega^2}{2} y_j^2 + V_{\ell}(x_j)\right]
 + \\ +
 \frac{C_\text{int}}{2} \sum_{i \neq j}
 \left[ (x_i - x_j)^2 + (y_i - y_j)^2 \right]^{-\alpha/2} \;,
\end{multline}
where the position and canonically conjugated momentum of atom $j$ in the plane are $(x_j,y_j)$ and $(p_{x,j},p_{y,j})$, respectively, while $V_{\ell}(x)$ is a weak confinement along the longitudinal direction, whose shape is conveniently chosen in order to fix the typical interparticle distance. The atom-atom interaction is proportional to the coupling constant $C_\text{int}$, which determines the strength of the interaction, and scales with the interparticle distance $r$ like $\sim 1/r^{\alpha}$, where the exponent $\alpha$ characterizes the nature of the atomic interaction. To provide some examples, for a system of ions we have $\alpha = 1$ and $C_\text{int} = Q^2 / 4 \pi \varepsilon_0$ (with $Q$ charge and $\varepsilon_0$ the vacuum permittivity), while for transversally pinned dipoles~\cite{Kollath,Morigi:Dipoli, Altman} it is  $\alpha = 3$, or even $\alpha = 6$ for Rydberg atoms in the induced dipole-dipole interaction regime~\cite{Pohl}. We remark that in general, one should check the conditions under 
which the effects of quantum degeneracy on the phase of the gas can be discarded. For atomic ions in typical experimental regimes the particles can be safely considered as distinguishable \cite{Javanainen}. 

In the following, starting from the Hamiltonian of Eq. \eqref{eq:original}, we review and critically discuss the basic steps of the mapping onto a lattice $\phi^4$ model. The latter is the basis of the numerical DMRG program, which is described in Sec. \ref{sec:simsim}.

\subsection{Longitudinal-transverse decoupling\\and natural units}

Although the longitudinal and the transverse motion of the atoms are coupled by the dynamics governed by the Hamiltonian in Eq. \eqref{eq:original}, it was argued in Ref.~\cite{Morigi:G.5} that the longitudinal dynamics plays a minor role in the structural transition and can be treated perturbatively. At lowest order in a gradient expansion, it was shown there that the zigzag mode is the soft mode of the transition, which is subject to a Landau-type potential (that possesses a Glodstone mode in three dimensions \cite{Morigi:G.5}). The mapping to a $\phi^4$ model has been explicitly derived in Refs. \cite{DeChiara,Shimshoni:2011a,Shimshoni:2011b}, by assuming that the coupling between longitudinal and transverse vibrations can be neglected sufficiently close to criticality.  

Following the arguments reported in Ref. \cite{Shimshoni:2011a}, we employ a model where only the transverse motion of the ions is included, namely
\begin{multline} \label{eq:transverse}
 H = \sum_{j = 1}^{L}
 \left[ \frac{p_{j}^2}{2M} + \frac{M \omega^2}{2} y_j^2 \right]
 + \\ +
 \frac{C_\text{int}}{2} \sum_{i \neq j}
 \left[ a^2 (i-j)^2 + (y_i - y_j)^2 \right]^{-\alpha/2} \;,
\end{multline}
where $p_j\equiv p_{y,j}$ and $a$ is the lattice constant. To recast the problem in dimensionless units, we adopt the lattice constant $a$ as the natural length scale and $\mathcal{E}_0 = {C_\text{int}} / a^{\alpha}$ as the energy scale. Then we rescale quantities as follows: $\tilde{y}_j = y_j / a$, $\tilde{H} = H / \mathcal{E}_0$, $\tilde{p}_j = p_{y,j} / \sqrt{ M \mathcal{E}_0}$, and finally $\tilde{\omega} = \omega / \sqrt{ \mathcal{E}_0 / M a^2}$. The Hamiltonian \eqref{eq:transverse} is thus rewritten as
\begin{equation} \label{eq:natural}
 H \!= \!\frac{1}{2}\!  \sum_{j = 1}^{L}\left(
 \tilde{p}_j^2 + \tilde{\omega}^2 \tilde{y}_j^2
 +
 \sum_{i \neq j}
 \frac{1}{
 \left[ (i-j)^2 + (\tilde{y}_i - \tilde{y}_j)^2 \right]^{\alpha / 2}
 }
 \right)\,,
\end{equation}
where the rescaled transverse trap frequency $\tilde{\omega}$ appears explicitly in the expression and is one of the two residual effective parameters (aside from $\alpha$). The other parameter appears in the commutation relation between the displacement $\tilde{y}$ and the transverse momentum $\tilde{p}$:
\begin{equation} 
 [\tilde{y}_i, \tilde{p}_j] = \imath g \delta_{i,j}\,,
\end{equation}
and reads
\begin{equation} \label{eq:gidef}
g = \sqrt{\frac{\hbar^2}{M a^2 \mathcal{E}_0}} = \hbar \,\sqrt{\frac{a^{\alpha - 2}}{M\, C_\text{int}}}.
\end{equation}
The parameter $g$ is dimensionless and corresponds to the square root of the ratio between the kinetic and the interaction energies. It 
plays an analogous role as the Planck constant in the rescaled commutator expression and thus provides a rough estimate of the impact of quantum fluctuations on transverse ordering. For this reason we refer to it as \emph{effective Planck constant}.

Typical values of $g$ depend on the experimental architecture and on the intrinsic nature of the atomic interaction. In particular, $g \propto a^{\alpha/2-1}$. For ions, $g$ increases with the density and takes values in the range between $10^{-5}$ and $10^{-4}$ (for $a \sim 1-10 \mu$m). On the other hand, for $\alpha>2$, which is the case for dipolar gases and Rydberg-dressed gases, $g$ increases as the density decreases, taking values $g > 10^{-2}$, whereas the lattice constant $a$ now spans from fractions of $\mu$m up to a few $\mu$m \cite{Pohl, Morigi:Dipoli, Altman}. Here, the quantum behavior becomes relevant at high densities. 

\subsection{Low-energy model}

Close to the transition point the critical behavior is determined by transverse fluctuations whose size is much smaller than the interparticle distance, namely, the lattice constant. This limit corresponds to the inequality $\sqrt{\langle\tilde{y}_j^2}\rangle\ll 1$, where the expectation value is taken over the ground state of the crystal. This condition allows one to substantially simplify the interaction term appearing in Eq. \eqref{eq:natural} by expanding the potential in powers of $(\tilde{y}_j - \tilde{y}_i)$. Taylor-expanding the interaction gives \cite{Morigi:G.5}
\begin{equation} \label{eq:anyord}
\begin{aligned}
 \frac{1}{\left[  x_0^2 + \delta \tilde{y}^2 \right]^{\alpha / 2} } &=
 \sum_{q = 0}^{t-1} \frac{\prod_{r=0}^{q-1} (\alpha + 2 r)}{\left(-2 \right)^q \; q!}
 \frac{\delta \tilde{y}^{2q}}{|x_0|^{\alpha + 2q}}
 + O[\delta \tilde{y}^{2t}] \\
 &= \sum_{q = 0}^{t-1} \frac{(-1)^{q} \, \Gamma(q + \textstyle{\frac{\alpha}{2}} )}{q! \;\; \Gamma( \textstyle{\frac{\alpha}{2}} )}
 \; \frac{\delta \tilde{y}^{2q}}{|x_0|^{\alpha + 2q}}
 + O[\delta \tilde{y}^{2t}] \\
\end{aligned}
\end{equation}
for arbitrary longitudinal separation $x_0 \geq 1$, where $\Gamma(\cdot)$ is the Euler Gamma function. Substituting this expression into equation \eqref{eq:natural} we obtain
\begin{multline} \label{eq:truexpand}
 \tilde{H} = E_0+\sum_{j = 1}^{L}
 \left[ \frac{\tilde{p}_j^2}{2} + \frac{\tilde{\omega}^2}{2} \tilde{y}_j^2 \right] \\ +
 \frac{1}{2} \sum_{q = 1}^{t-1} (-1)^q \; b_q(\alpha) \sum_{i \neq j}
 \frac{ (\tilde{y}_i - \tilde{y}_j)^{2q} }{ |i - j|^{\alpha+2q} }
 + O[\delta \tilde{y}^{2t}]
  \;.
\end{multline}
where $E_0$ is a constant and corresponds to the classical ground-state energy of the linear chain \cite{Dubin1997,Morigi:G.7},  while $b_q(\alpha) = \Gamma(q + {\frac{\alpha}{2}} ) / q! \;\Gamma( {\frac{\alpha}{2}} )$ is a positive coupling constant. The previous manipulation makes the problem easier to address by numerical means, since now the parameter $\alpha$ only enters in the coupling coefficients \cite{Shimshoni:2011a}.

We remark that it is important to truncate the expansion at an odd $t$ order, as we did in Eq. \eqref{eq:truexpand}. In fact, this guarantees that the truncated interaction potential in \eqref{eq:anyord} and \eqref{eq:truexpand} is ultimately bounded from below, which is mandatory for avoiding convergence/stability issues of any numerical method we wish to employ.

\subsection{Recasting into a short range theory}

Dealing with a long-range model as in Hamiltonian \eqref{eq:truexpand} is numerically demanding and cumbersome, especially with DMRG, where it leads to slower computational scaling with the system size. For this reason we will adopt an approximation that further simplifies the model. 

Based on the arguments introduced in Ref.~\cite{Morigi:G.5, DeChiara:2008}, in the Appendix we show the detailed derivation of a short-range model capable of mimicking quantitatively the linear-zigzag quantum phase transition of the Wigner crystal. We stress that this technique is not based on truncating the interactions at a finite distance:  in fact, due to the collective nature of the phononic mode driving the instability (soft mode), truncation would lead to a systematic error in determining the phase diagram. On the contrary, the mapping we adopt reproduces the multi-phonon dispersion bands around the soft mode as faithfully as possible with a nearest-neighbor interacting theory. The starting point is the assumption that at sufficiently low energies, the soft mode, which has quasimomentum $k_0 =\pi$ (in natural lattice units where $a = 1$),
interacts primarily with its neighboring modes in the Brillouin zone $[-\pi,\pi]$. Therefore, for any expansion term $q$ in
Eq.~\eqref{eq:truexpand} we construct the corresponding $q \to q$ phonon scattering function $\Xi[k]$, and approximate it with a short-range interaction matching up to the second order in $\delta k = (k - k_0)$ around $k_0 = \pi$. The algebraic technique employed to achieve this is detailed in appendix \ref{sec:append}. Here we simply report the resulting low-energy Hamiltonian \cite{Shimshoni:2011b}:
\begin{multline} \label{eq:serious}
 \tilde{H} \simeq \frac{1}{2} \sum_{j = 1}^{L}
 \left[ \tilde{p}_j^2 + \tilde{\omega}^2 \tilde{y}_j^2 + \sum_{q = 1}^{t-1} (-1)^q \times \right.
 \\
 \left. \times \left( \eM_q(\alpha) \,\tilde{y}_j^{2q} - \mathcal{N}_q(\alpha) \left( \tilde{y}_j^{q} - (- \tilde{y}_{j+1} )^{q} \right)^2
 \right) \right]
  \;.
\end{multline}
The on-site fields $\eM$ and coupling constants $\eN$ are now functions solely of $q$ and $\alpha$, and not of other physical parameters. They read respectively as follows:
\begin{equation} \label{eq:onsitefield}
 \eM_{q} =
 \frac{\left(2^{\alpha+2q} - 1 \right) \Gamma(q + \textstyle{\frac{\alpha}{2}})}{
 q! \;2^{\alpha - 1} \; \Gamma(\textstyle{\frac{\alpha}{2}})}\, \zeta(\alpha+2q)\,,
\end{equation}
\begin{equation} \label{eq:offsitefield}
 \begin{aligned}
 \eN_{1} &=
 \left\{ \begin{aligned}
  &\ln 2 \qquad &\mbox{for } \alpha = 1 &\\
  &\frac{2^\alpha - 2}{2^{\alpha}}\;\alpha \zeta(\alpha) \qquad &\mbox{for } \alpha > 1\,,&
 \end{aligned} \right. \\[7pt]
 \eN_{q > 1} &=\!
 \frac{2q-1}{q}\;	
 \frac{\left(2^{\alpha+2q-2} - 1 \right) \Gamma(q + \textstyle{\frac{\alpha}{2}})}{
 q! \; 2^{\alpha - 1} \; \Gamma(\textstyle{\frac{\alpha}{2}})}
 \zeta(\alpha+2Q-2)\,.
 \end{aligned}
\end{equation}
Let us remark that these couplings capture the collective character of the excitation modes, whose signature is the presence of the Riemann zeta function $\zeta(\cdot)$ in their expression.

We stress that the resulting short-range model \eqref{eq:serious} successfully mimics the dynamics of the original model \eqref{eq:truexpand} only when the excitation energies we are dealing with are small compared with the energy scale associated with the phononic bandwidth. When this is the case, then the modes with quasimomentum $k$ far from $k_0$ (such that $1\lesssim |\delta k| \leq \pi$) play a negligible role in the description of the critical behavior. Since we are interested in the ground state properties, this requirement is satisfied by definition, and thus we can safely accept the approximation.

Moreover, we point out that coefficients $\eM_q$ and $\eN_q$ grow for large $q$ roughly as $\sim 4^q \; q^{1- \alpha/2}$: this means that the series in Eq.~\eqref{eq:serious} converges only close enough to the linear phase, i.e., $| \langle \tilde{y}^{q} \rangle | < 2^{-q}$. Again, since we are investigating the critical behavior, this requirement is easily achieved.
Let us finally remark that  the present
formulation given by Eqs.~\eqref{eq:serious}, \eqref{eq:onsitefield} and \eqref{eq:offsitefield}
of the long-range into short-range mapping was first shown in Refs.~\cite{Morigi:G.5, DeChiara:2008}.

\subsection{Fourth-order expansion}

In this work, we keep up to four total expansion orders in the series described in Eq. \eqref{eq:serious}: we include in the picture the $\tilde{y}_j^4$ local term, as well as the $(\tilde{y}_j + \tilde{y}_{j+1})^2$ interaction, which is second order in the displacement and second order in $\delta k$. We have checked that neglecting further expansion orders generates errors that are compatible with or smaller than errors due to other aspects of the numerical technique we employ afterwards.

The Hamiltonian we simulate, as a function of all the residual parameters, reads
\begin{multline} \label{eq:actualmodel}
 \tilde{H}(\alpha, g, \tilde{\omega}) = \frac{1}{2} \sum_{j = 1}^{L}
 \left[ - g^2 \frac{\partial^2}{\partial \tilde{y}_j^2} + \left( \tilde{\omega}^2 - \eM_1(\alpha) \right) \tilde{y}_j^2 +
 \vphantom{\sum} \right.
 \\
 \left. +\, \eN_1(\alpha) \left( \tilde{y}_j + \tilde{y}_{j+1} \right)^2 + \eM_2(\alpha)\; \tilde{y}_j^4 \right],
\end{multline}
where we already substituted $\tilde{p}_j = - i g(\partial/\partial \tilde{y}_j)$
to make the parameter $g$ appear explicitly in the expression.
Equation~\eqref{eq:actualmodel} reveals an important physical aspect of the linear-zigzag transition: As long as the coupling with the axial vibrations can be neglected (or just give rise to a renormalization of the parameters of the transverse Hamiltonian), the critical behavior at the phase transition does not depend on the interaction-range scaling exponent $\alpha$. In fact,
given two values $\alpha$ and $\alpha'$
it is possible to map Hamiltonian
$\tilde{H}_0 = \tilde{H}(\alpha, g, \tilde{\omega})$ into
$\tilde{H}' = \tilde{H}(\alpha', g', \tilde{\omega}')$,
by a simple rescaling of energies and length scales \cite{Shimshoni:2011b}.
Precisely, by requiring $\tilde{y}_j' = u \tilde{y}_j$ and $\tilde{H}'= v \tilde{H}_0$ we obtain
\begin{equation}
 u = \sqrt{\frac{\eN_1 (\alpha') \; \eM_2 (\alpha)}{\eN_1 (\alpha) \; \eM_2 (\alpha')}}
 \qquad
 v = \frac{\eN_1^2 (\alpha') \; \eM_2 (\alpha)}{\eN_1^2 (\alpha) \; \eM_2 (\alpha')}.
\end{equation}
while the other parameters $g$ and $\tilde{\omega}$ must transform as
\begin{align} \label{eq:paramconnect}
 \tilde{\omega}'^2
 &= \eM_1(\alpha') + \frac{\eN_1(\alpha')}{\eN_1(\alpha)}\left( \tilde{\omega}^2 - \eM_1(\alpha) \right)
 \\  \label{eq:paradue}
 g' &= g \cdot
 \frac{\eM_2 (\alpha)}{\eM_2 (\alpha')}
 \left( \frac{\eN_1 (\alpha')}{\eN_1 (\alpha)} \right)^{\frac{3}{2}}.
\end{align}
Rephrasing, all the linear-zigzag physics formulated according to Eq. \eqref{eq:actualmodel}, for various values of $\alpha$,
are equivalent: the critical behaviour shows the same properties and the phase diagrams in the external parameters space ($g$ and $\tilde{\omega}$) transform into one another according to relations \eqref{eq:paramconnect} and \eqref{eq:paradue}.
We remark that this argument is valid as long as one can safely decouple the axial from the transverse motion, which appears correct for ion Coulomb chains.

We will from now on drop the functional dependence of coefficients $\eM$ and $\eN$ on $\alpha$. Whenever a specific value of $\alpha$ is implicitly assumed, we will be referring to the ion Wigner crystal scenario \ie $\alpha = 1$. In this setup the coefficients read $\eM_1 = 7 \,\zeta(3) / 2 \simeq 4.2072$, $\eN_1 = \ln 2 \simeq 0.6931$ and $\eM_2 = 93 \,\zeta(5) / 8 \simeq 12.0543$.

\subsection{Connection with the $\phi^4$ model}

Remarkably, the effective model in Eq. \eqref{eq:actualmodel} is closely related to a $\phi^4$ field
theory~\cite{Morigi:G.5,DeChiara,Shimshoni:2011b}: it is basically
an antiferromagnetic formulation of the $\phi^4$ theory on a lattice.
The typical formalism in field theory, where a real
scalar massive field $\phi(x,t)$, undergoing a Klein-Gordon motion, is dressed by a pointwise interaction
of the form $\phi^4$, reads
\begin{equation} \label{eq:phifour}
 \eL_{\phi^4} = \int \left[ \frac{1}{2} \; \partial^{\mu} \phi \, \partial_{\mu} \phi \; -
 \frac{m^2}{2} \, \phi^2 - \frac{\lambda}{4!} \phi^4 \right] dx,
\end{equation}
with flat space-time metric:
$\partial^{\mu} \phi \, \partial_{\mu} \phi = ( \partial_t \phi )^2 - ( \partial_x \phi )^2$,
in units of $\hbar = c = 1$.
We now briefly summarize the steps showing that a Lagrangian of the type \eqref{eq:phifour} can be obtained from
Eq. \eqref{eq:actualmodel}
by means of three simple steps: field staggerization, continuum limit, and canonical rescaling.
Precisely, let $\phi_j = (-1)^j \tilde{y}_j$ be the scalar quantum field. Now, going from a lattice to continuous space
yields
\begin{multline} \label{eq:contimodel}
 \tilde{H} = \frac{1}{2} \int
 \left[ g^2 \pi^2(x) + \left( \tilde{\omega}^2 - \eM_1 \right) \phi^2(x) + \right.
 \\
 \left. +\, \eN_1 \left( \partial_x \phi \right)^2 + \eM_2\; \phi^4(x) \right] dx,
\end{multline}
where we performed the substitution $(\phi_{j+1} - \phi_j) \to \partial_x \phi$,
and $\pi(x)$ is the canonically-conjugated field, fulfilling the commutation relation at equal times $[\phi(x),\pi(x')]=i\delta(x-x')$.
In order to obtain an equation of type \eqref{eq:phifour}, we need to rescale
energies and fields ($\tilde{H} \to \tilde{H} / g \sqrt{\eN_1}$, $\phi \to \phi\, (g^2 / N_1)^{1/4}$), followed by a standard Legendre transformation. This leads to the Lagrangian
\begin{equation} \label{eq:lege}
 \eL \!=\! \frac{1}{2}\! \int\! \left[ ( \partial_t \phi )^2 - ( \partial_x \phi )^2 -
 \frac{\tilde{\omega}^2 \!-\! \eM_1}{\eN_1} \, \phi^2 -
 \frac{g \eM_2}{\eN_1^{3/2}} \phi^4 \right] dx,
\end{equation}
which connects to \eqref{eq:phifour} via the relations
\begin{equation}
 m = \sqrt{\frac{\omega^2 - \eM_1}{\eN_1}}
 \quad \mbox{and} \quad
 \lambda = 12 \,\frac{g \,\eM_2}{\eN_1^{3/2}}.
\end{equation}
The one-dimensional $\phi^4$ field theory on a lattice has been already addressed by means of numerical simulation, both by Montecarlo methods \cite{Barma, MontecarloP4} and also by DMRG  \cite{Sugihara}. In this paper we provide a complete, exhaustive characterization of its quantum criticality,
while exploring the whole phase-diagram boundary, therefore extending and complementing the results we presented in Ref. \cite{NostroANDP}.

\section{DMRG Simulation Details}\label{sec:simsim}

The Density Matrix Renormalization Group (DMRG) is a method developed in the early 90's \cite{White92} which has proven successful for a large variety of one-dimensional many-body quantum problems \cite{PrimoMPS}. The key to its success relies on the fact that the entanglement description capabilities of DMRG match exactly the typical entanglement scaling laws of ground states in one-dimensional quantum systems (and low-lying energy states too) \cite{Arealaws}.

Simulating the effective lattice staggered $\phi^4$ model of Eq.~\eqref{eq:actualmodel} with DMRG is feasible,
but it still requires some additional careful numerical treatment\cite{Localtrunc}.
We will discuss these expedients in the present section.

\begin{figure}
 \begin{flushright}
 \begin{overpic}[width = 230pt, unit=1pt]{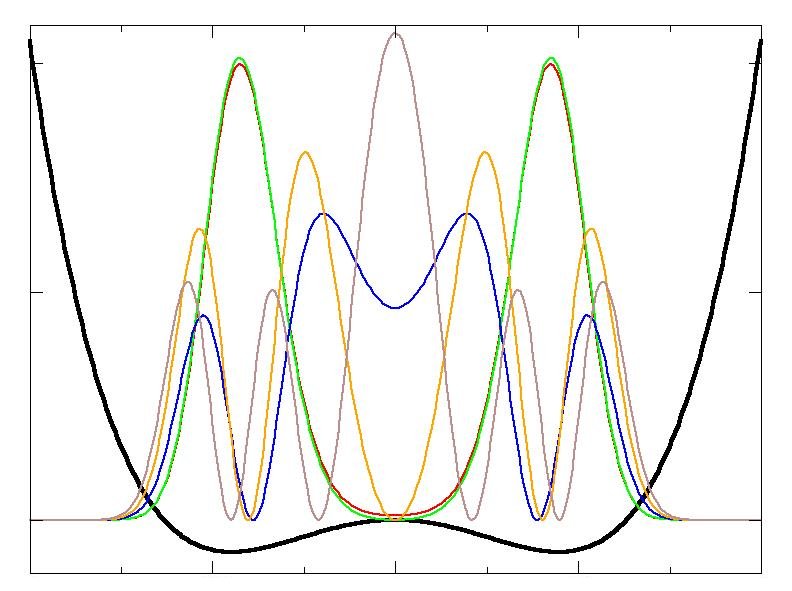}
  \put(117, -6){$\tilde{y}$}
  \put(-12, 128){$|\psi_q|^2$}
 \end{overpic}
 \end{flushright}
\caption{ \label{fig:lobas}
Probability densities $|\psi_q(\tilde{y})|^2$ as a function of $\tilde{y}$ for the lowest five quantum energy eigenstates
of the local Hamiltonian of Eq.~\eqref{eq:localham}: $q = 1$ (red line), $q=2$ (green line), $q = 3$ (blue line),
$q = 4$ (yellow line) and $q = 5$ (grey line).
For easy reference we also plot the double well potential profile (black thick line) of the Hamiltonian,
in arbitrary energyscale units.
}
\end{figure}

\subsection{Local basis selection}

Traditional DMRG architectures\cite{White92}
are tailored to models where the local space is a finite-dimensional Hilbert space, like for instance for spin models. In the scenario here considered, however,
the local space is a continuous quantum variable, with the usual Lie algebra $\{\tilde{y}, \tilde{p}\}$ of the quantum particle motion.
To circumvent this obstacle, we adopt a local space truncation approach that was thoroughly discussed in Ref.~\cite{Localtrunc}.
Namely, 
we define a related single-particle quantum problem $H_{\text{loc}}$,
then find its $d$ lowest-energy eigenfunctions $|\psi_q\rangle$ and adopt them as local basis
$\{|\psi_q\rangle \}_{q = 1..d}$ for the many-body problem.
If the many-body Hamiltonian with nearest-neighbor interaction is
$\tilde{H} = \sum_ j H_{\text{loc}}^{(j)} + H_{\text{int}}^{(j,j+1)}$, then
the simulation is more accurate for a given $d$ (or it requires smaller $d$ to achieve some target precision),
when the interaction energy $\langle H_{\text{int}} \rangle$ is smaller in modulus; that is, when $H_{\text{int}}$
can be treated as a perturbation.

For the case under study we argue that considering the whole $(\tilde{y}_j + \tilde{y}_{j+1})^2$ term as
the interaction part is a more perturbative approach than just taking the double product $2 \tilde{y}_j \tilde{y}_{j+1}$.
Indeed, while in the linear phase the expectation value on the ground state of
two terms is of roughly the same magnitude, in the zigzag phase the first
is definitely closer to zero, and thus more appropriate to be chosen as interaction part.
According to this scheme, the local Hamiltonian reads
\begin{equation} \label{eq:localham}
 H_{\text{loc}} = \frac{1}{2}
 \left[ - g^2 \frac{\partial^2}{\partial \tilde{y}^2} + \left( \tilde{\omega}^2 - \eM_1\right) \tilde{y}^2 +
 \eM_2\; \tilde{y}^4 \right],
\end{equation}
describing the motion of a quantum particle in a harmonic-quartic potential. Thanks to the translational invariance
of Eq.~\eqref{eq:actualmodel}, we just have to solve the problem
\eqref{eq:localham} once per given $g$ and $\tilde{\omega}$, and use the resulting basis for every site.
In order to find the low-energy eigenstates of Eq.~\eqref{eq:localham},
we solve $H_{\text{loc}} |\psi_q\rangle = E_q |\psi_q\rangle$ exactly by means of linear algebra numerical methods
to diagonalize tridiagonal matrices.
Afterwards, we express the single-body computational basis as $|q_j\rangle \equiv |\psi_{q}\rangle$, which corresponds to the
atom at site $j$ being in the orbital state $|\psi_q\rangle$.
The resulting many-body computational basis is made of tensor product states of the single-site basis states
$|q_1\,q_2\ldots q_L\rangle = |q_1\rangle \otimes |q_2\rangle \otimes \ldots \otimes |q_L \rangle$.
In this formalism, the global hamiltonian then reads
\begin{equation} \label{eq:dmrgready}
 \tilde{H} = \sum_j ( A_j + \eN_1 \,W_j ) + \eN_1 \,Y_j \otimes Y_{j+1}
\end{equation}
where the matrices $\Theta_j$ ($\Theta = A,W,Y$) are single-site operators acting on site $j$, although their
matrix elements do not explicitely depend on $j$ thanks to translational invariance. Specifically, we have
$A_j = \sum_{q = 1}^d E_q |q_j\rangle \langle q_j|$, then
$Y_j = \sum_{q,q'}^{d} |q_j\rangle \langle \psi_q | \,\tilde{y}\, |\psi_{q'}\rangle \langle q'_j|$ and
$W_j = \sum_{q,q'}^{d} |q_j\rangle \langle \psi_q | \,\tilde{y}^2\, |\psi_{q'}\rangle \langle q'_j|$.
We explicitly express the change of basis $|\psi_q \rangle\langle q_j|$ to stress the fact that we are projecting the
space over the first $d$ local states. 
Notice that, as a consequence of this, $W \neq Y^2$.

%
\begin{figure}
 \begin{center}
 \begin{overpic}[width = \columnwidth, unit=1pt]{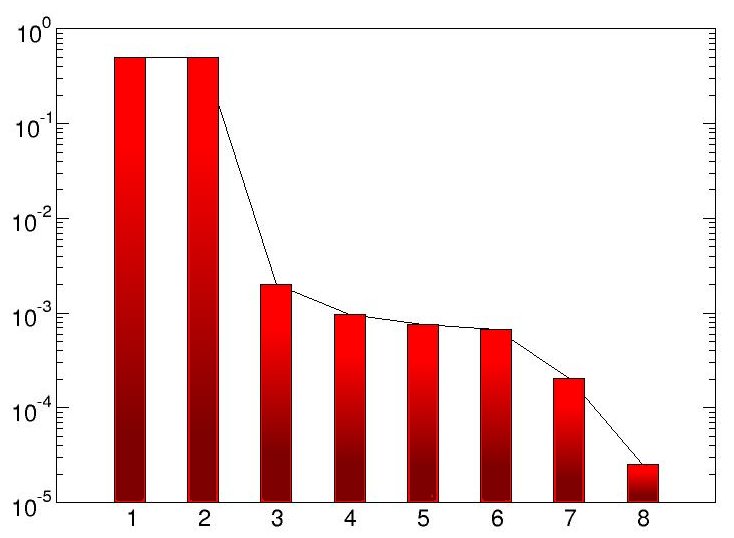}
  \put(106, -8){Level index $q$}
  \put(123, 86){$p(q)$}
 \end{overpic}
 \end{center}
\caption{ \label{fig:histo}
Probability distribution $p(q)$, plotted as a function of the local basis number $q$.
$p(q)$ is defined as $p(q) = \langle q | \varrho_j | q \rangle$ where
$\varrho_j = \trace_{j'\neq j}\{|\Psi^{L}_{g,\tilde{\omega}}\rangle\langle\Psi^{L}_{g,\tilde{\omega}}|\}$
is the reduced density matrix of an arbitrary site $j$ in the bulk
of the many-body ground state $|\Psi^{L}_{g,\tilde{\omega}}\rangle$,
simulated by DMRG. Here the system is in the zigzag phase, but close to critical point ($g = 0.03$, $\tilde{\omega}^2 = 1.0$):
The first two levels (symmetric and antisymmetric double peaks respectively)
have equal populations. The third level probability is
two orders of magnitude smaller, and rapidly decaying.
}
\end{figure}

Typical solutions of the local problem defined by Eq.~\eqref{eq:localham}, in the deep quantum regime ($g = 0.1$),
are displayed in Fig.~\ref{fig:lobas}.
Both the second  and fourth excited states (blue and grey line, respectively) show a non-negligible probability density at the barrier point ($\tilde{y} = 0$). This reveals that the quantum fluctuations enhance substantial tunneling between the two potential wells, thus ultimately making the linear (disordered) phase energetically favourable.

In order to keep track of the error generated by truncating the basis $|\psi_q\rangle$ to a dimension $d$, we performed several calculations of the same problem (under identical environment parameters) for various values $d \sim 2 \ldots 100$, until we located convergence of the outcomes.
Furthermore, we kept track of the populations of various basis levels on every site: namely
we verify that the occupation probabilities in the one-site reduced density matrix
decrease roughly exponentially with the level index $q$.
This provides a meaningful lower bound for the error generated by the truncation. Figure~\ref{fig:histo} displays the populations $p(q)$ of the first eight local basis levels, obtained after the numerical simulation of the ground state for a given set of parameters (further levels are of order of magnitude below $10^{-5}$ and not shown in the plot). In all the cases we considered, the populations $p(q)$
decay very rapidly with the level index $q$. They can usually be bounded from above by an exponential
decay $p(q \geq 3) \leq e^{-\Lambda q}$. For the case in the figure, $\Lambda \simeq 1.7$.

Hamiltonian \eqref{eq:dmrgready} is ready for simulation, and via standard DMRG architecture \cite{White92} we searched for its ground state for finite system size $L$ with Open Boundary Conditions (OBC). The latter choice is due to a natural tendency of DMRG with respect to OBC: in this scenario it converges faster, and it has enhanced precision and stability.
In the various physical systems considered, a local basis dimension $d \sim 30$ and a
DMRG bondlink $D \sim 50$ were sufficient to make the results converge permanently under our precision
(typically $10^{-10}$ of relative numerical precision).

\subsection{Measured order parameter}

A final remark regards the identification of the phase across the transition. A drawback of working on finite-size quantum samples is the impossibility of achieving a truly spontaneous symmetry-broken phase. In our problem, the two possible zigzag configurations bear a nonzero interference term, which encourages the even superposition of the two as true ground state, thus restoring the $\mathbb{Z}_2$ parity symmetry. A standard technique known in literature\cite{Mussardo}
to circumvent this issue is employing structure factor-based order parameters, which are insensitive to symmetry breaking. We briefly review this strategy for easy reference to the reader. The order parameter we choose is the
\emph{square root of the structure factor density} calculated at the soft mode $k_0 = \pi$, precisely
\begin{equation} \label{eq:strufa}
 \xi_L(g, \tilde{\omega}) = \sqrt{\frac{1}{L^2} \sum_{j,j'}^L (-1)^{j-j'} \langle \Psi_{g, \tilde{\omega}}^L |
 Y_j \otimes Y_{j'} | \Psi^L_{g, \tilde{\omega}} \rangle
 }\;,
\end{equation}
where $ | \Psi_{g, \tilde{\omega}}^L \rangle$ is the many-body ground state calculated via DMRG under parameters
$g$, $\tilde{\omega}$ and size $L$.
This is clearly a non-extensive quantity, and it can be shown to exactly coincide with the standard antiferromagnetic
order parameter $\bar{m} = L^{-1} \sum_j^L (-1)^j \langle Y_j \rangle$ in the thermodynamical limit.
Basically if we assume that correlations can be split into a classical and a quantum contribution, \ie
$\langle Y_j Y_{j'} \rangle \simeq \langle Y_j \rangle \langle Y_{j'} \rangle + f_q(j-j')$ respectively,
then the quantum part becomes irrelevant when evaluating $\xi_L$.
In fact, it is either $f_q(j-j') \sim |j-j'|^{-\nu}$ (critical scenario, with $\nu > 0$), or
$f_q(j-j') \sim e^{-|j-j'|/\lambda}$ (noncritical scenario). In both cases it holds
\begin{equation}
 \left|  \sum_{j,j'}^L \frac{(-1)^{(j-j')}}{L^2} f_q(j-j') \right|
 \leq \frac{ f_q(1) + \int_1^{L} \left| f_q(x) \right| dx}{L}  \to 0\,.
\end{equation}
Consequently, we obtain
\begin{equation} \label{eq:stru2}
 \xi_L = \sqrt{\sum_{j,j'}^L (-1)^{(j-j')}
 \frac{ \langle Y_j \rangle \langle Y_{j'}  \rangle}{L^2} } \simeq \sqrt{\bar{m}^2} = \bar{m}\,,
\end{equation}
which tells us that $\xi_L$ has all the properties of a local antiferromagnetization without suffering from finite-size symmetry breaking issues, since it is based on two-point correlation measurements and not on local observations.

\section{Phase Diagram}\label{sec:phasedia}

\begin{figure}
 \begin{center}
 \begin{overpic}[width = \columnwidth, unit=1pt]{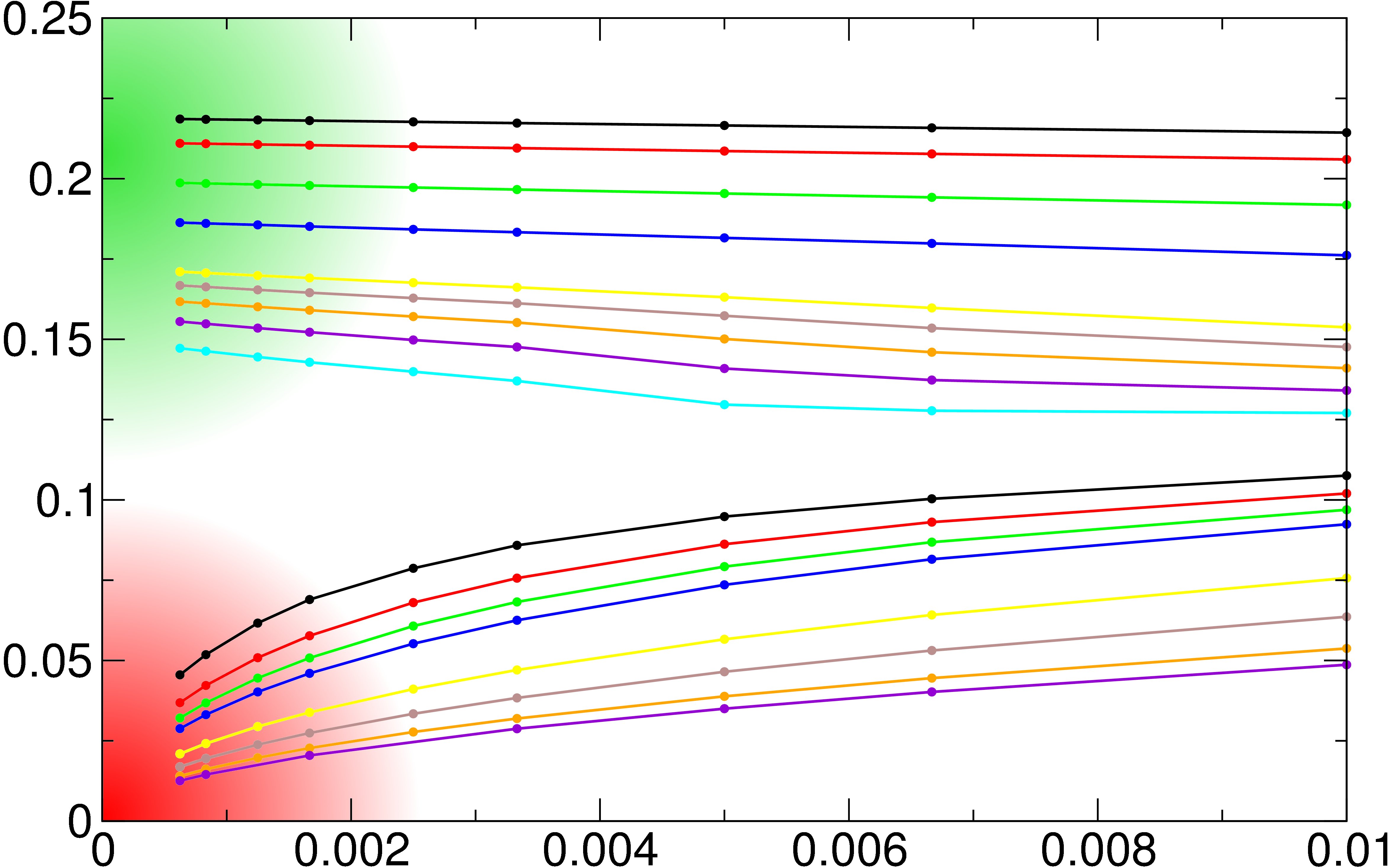}
  \put(0, 78){$\xi_L$}
  \put(121, -10){$1/L$}
 \end{overpic}
 \end{center}
\caption{ \label{fig:TDPhase} (color online)
 Zigzag-order parameter $\xi_L(g, \tilde{\omega})$ as a function of $1/L$.
 Here the lattice size $L$ ranges within $100 \leq L \leq 1600$, while
 other parameters are $g = 0.08$ and $d = 14$.
 The points have been numerically evaluated, the lines
 are eye-guides.
 Each set of data corresponds to a different value of the square trap frequency
 $\tilde{\omega}^2 = 1.30, 1.32, 1.34, \ldots ,1.58, 1.60$ (from top to bottom).
 The zigzag order parameter is obtained via DMRG using the square root of structure-factor density
 of Eq. \eqref{eq:strufa}.
 }
\end{figure}

\begin{figure}
 \begin{center}
 \begin{overpic}[width = \columnwidth, unit=1pt]{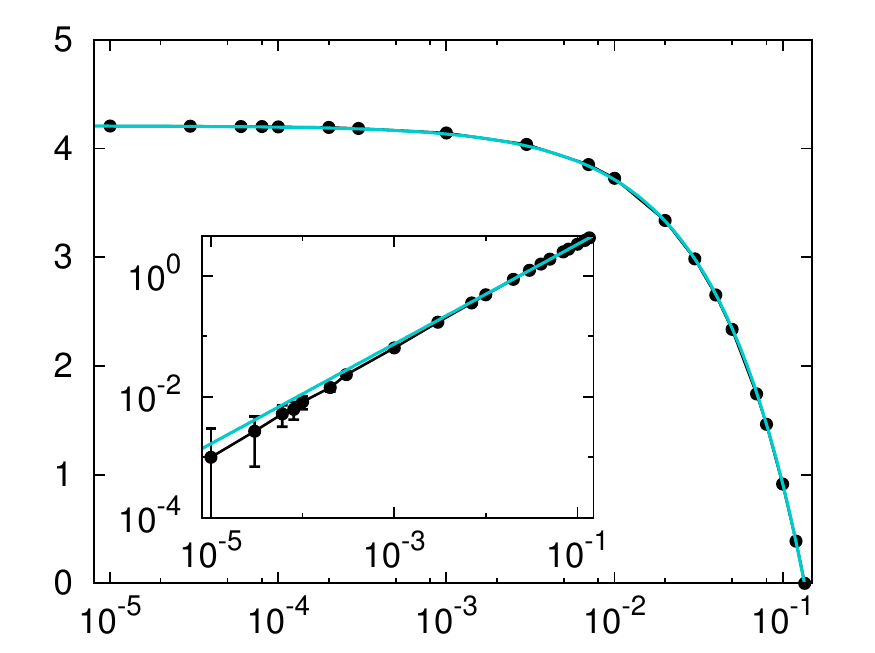}
  \put(-3, 92){\large $\tilde{\omega}_c^2(g)$}
  \put(194, 0){\large $g$}
  \put(34, 122){$|\Delta \tilde{\omega}_c^2 |$}
  \put(170, 152){Linear phase}
  \put(85, 127){Zigzag phase}
  \put(70, 100){Zigzag}
  \put(120, 60){Linear}
 \end{overpic}
 \end{center}
\caption{ \label{fig:phasedia} (color online)
Phase diagram of an array of ions (\ie $\alpha = 1$) in the $(g, \tilde{\omega}^2)$ parameter space,
at the thermodynamical limit.
Inset: displacement of the critical square frequency $|\Delta \tilde{\omega}^2_c| $ from the classical value,
as a function of $g$. Cyan lines represent a power-law fit of the whole data curve
using $|\Delta \tilde{\omega}^2_c| (g)= u \,g^{v}$ via $u$ and $v$
(see text).}
\end{figure}
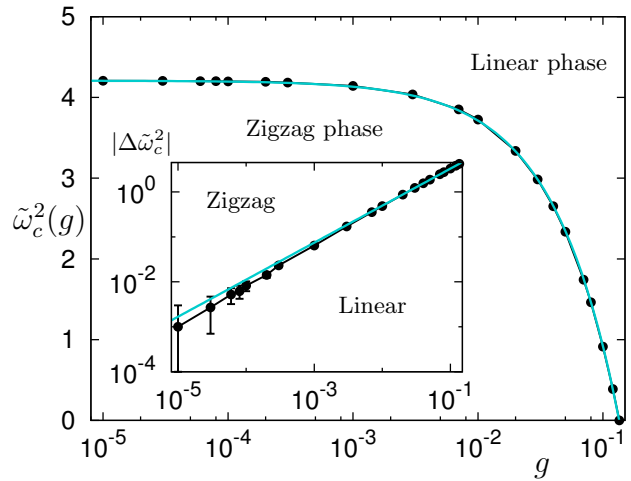

Some of the results presented in this and the following sections were previously reported in Ref.~\cite{NostroANDP},
in particular Figs.~\ref{fig:phasedia}, \ref{fig:finsiz}, \ref{fig:central} that we report for completeness and comfort to the reader.
Here we describe the complete numerical derivation which allows to process these physical quantities from raw simulation data,
and we give additional comments which clarify technical issues about those results.

First of all,
we characterize the phase of the ground state for a given point in the parameters space ($g, \tilde{\omega}^2$). More precisely, we simulate the same OBC problem for various system sizes $L$, typically up to 3000 sites, with the prescriptions detailed in the previous section. For each simulation, we record the zigzag order parameter $\xi_L(g, \tilde{\omega})$ introduced in Eq. \eqref{eq:strufa} by measuring every two-point correlator. Finally, we fit the thermodynamic limit $\xi_{\infty} = \lim_{L \to \infty} \xi_L$ and we discriminate whether its value is zero, which detects the linear phase, or it is nonzero, thus revealing the zigzag phase. Figure~\ref{fig:TDPhase} displays the order parameter $\xi_L$ as a function of the length $L$ of the sample.
Various data sets are plotted, each one corresponding to a different value of $\tilde{\omega}^2$ (at the same
value of $g$).
Every curve $\xi(L)$ is fitted via various fit functions, polynomial in the inverse length, 
and one clearly
sees that the order parameter is typically a very smooth function of the chain size. 

Since we are interested in determining the phase diagram, we need to discriminate when $\xi_{\infty}$ is zero and when it is not, regardless of the function used to fit the value. This allows us to detect the phase boundary
in the parameter space ($g, \tilde{\omega}^2$) with an uncertainty 
of the order of $10^{-3}$. Figure~\ref{fig:phasedia} displays the phase diagram,
which has been derived after locating the critical trap frequency value $\tilde{\omega}_c(g)$  for several values of $g$
in the range $10^{-5} \leq g \leq 0.2$. As expected, when $g$ increases, the magnitude of quantum fluctuations increases and with it the range of the disordered phase. In accordance to this conjecture, $\tilde{\omega}_c(g)$ is a monotonically decreasing function of $g$.
We determine the shift of the critical point $\tilde{\omega}_c^2(g)$ with respect to the one predicted by the classical theory \cite{Morigi:G.5}, which is given by $\tilde{\omega}_c^2(0) = \eM _1$. This corresponds to the quantity $\Delta \tilde{\omega}_c^2(g)\equiv \eM _1 - \tilde{\omega}_c^2(g)$, and we obtain that it scales with $g$ according to a power-law behaviour given by 
$$\Delta \tilde{\omega}_c^2(g) = (21.91 \pm 0.01) \cdot g^{(0.823 \pm 0.003)}\,.$$

We remark that we also employed other data-processing strategies to draw the critical boundary, which will be detailed in the next section. These are
divergence of the correlation length, entanglement area-law violation,
and finite-size scaling. In all the cases these methods have proven compatible,
albeit less precise, to the structure-factor density technique.

\section{Critical exponents}\label{sec:uniclass}

The largest local symmetry group under which our model $\tilde{H}$ is invariant is parity symmetry. Specifically, let $R$ be the unitary  reflection operator along the transverse direction, \ie $R \:\tilde{y}\: R^{\dagger} = -\tilde{y}$. Then $R = R^{\dagger}$, $R^2 = 1$, which is a representation of the $\mathbb{Z}_2$ group, and it is straightforward to check that it is a symmetry of the Hamiltonian, i.e.~$[\tilde{H}, R^{\otimes L}] = 0$. In the thermodynamical limit, the zigzag phase has a two-fold degenerate ground space, which spontaneously breaks the symmetry $R^{\otimes L}$.
Starting from this symmetry argument, conformal field theory predicts that
the continuous model should exhibit the same universality class of the 1D quantum Ising model with transverse field \cite{Sachdev, Mussardo, Cardy}. We are now going to test the validity of this claim in the lattice model by numerically evaluating various critical exponents, and comparing them with the corresponding theoretical predictions.

\begin{figure}
 \begin{center}
 \begin{overpic}[width = \columnwidth, unit=1pt]{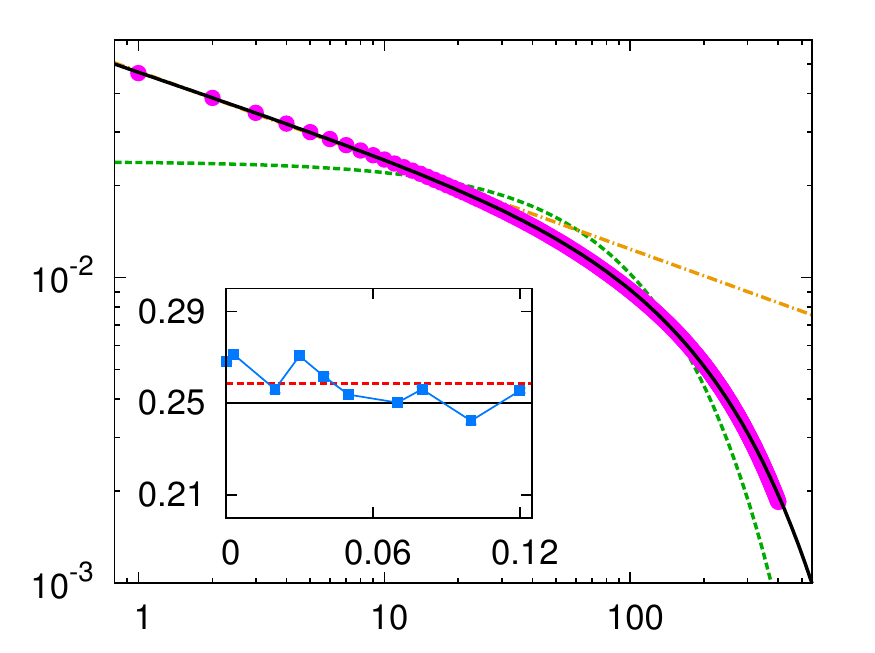}
  \put(-4, 134){\large $G(\Delta j)$}
  \put(215, 2){\large $\Delta j$}
  \put(124, 28){$g$}
  \put(50, 110){$\eta$}
 \end{overpic}
 \end{center}
\caption{ \label{fig:correl} (color online)
Two point correlations $G(\Delta j)$ of the displacement $\tilde{y}$ as a function of the distance
$\Delta j$ (magenta dots), acquired via
numerical ground-state simulation with DMRG.
Data are fitted using
$G_{\rm fit}(\Delta j)$ in Eq. \eqref{G:fit} (black solid line).
The green and orange lines show
fits via an exponential and a power-law respectively.
Inset: fitted anomalous dimension exponents $\eta$ over several points of the critical boundary,
characterized by different values of $g$, the dashed line shows the average value
$\eta = 0.258 \pm 0.012$, the solid line is the value of $1/4$ predicted from the Ising model theory.
}
\end{figure}
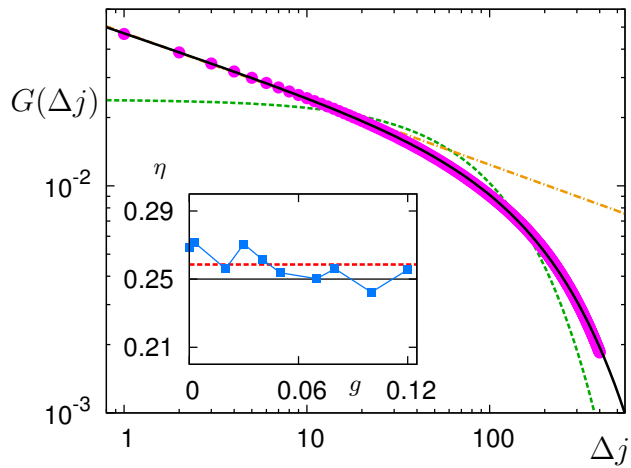

\subsection{Anomalous dimension $\eta$}

The transverse coordinate $\tilde{y}$ of the lattice field model plays the same role as the interaction Pauli matrix of the Ising model \cite{Shimshoni:2011a} (namely $\sigma^z$ if one writes $H_{\text{Ising}} = \sum_j \sigma^z_j \sigma^z_{j+1} + B \sigma^x_j$). As a result, studying quantum correlation functions in the $\tilde{y}$ direction, such as
\begin{equation}
Q(j,j') \equiv \langle \tilde{y}_j \tilde{y}_{j'} \rangle - \langle \tilde{y}_j \rangle \langle \tilde{y}_{j'} \rangle\,,
\end{equation}
should reveal the \emph{anomalous dimension} critical exponent $\eta$ \cite{Sachdev}.
To make sure that the scaling of correlations depends only on the distance $\Delta j \equiv |j-j'|$ of the pair and is insensitive to boundary effects, we enlarge our system size up to thousands of sites, and average over the bulk:
the expression 
\begin{equation} \label{eq:bulky}
G(\Delta j) = (-1)^{\Delta j}\, \tilde{L}^{-1} \sum_{j}^{\tilde{L}} Q(j,j+\Delta j)
\end{equation} 
is chosen in such a way as to take into account staggerization as well.
We calculate $G(\Delta j)$ from the numerical data summing over $j$ sites in \eqref{eq:bulky} such that we disregard sites sitting too
close to the boundary. Namely, we include only pairs $\{j,j+\Delta j\}$ located within the central third of the chain.

Moreover, since we are never exactly simulating the critical point, we must introduce a finite-correlation length $\lambda$ and evaluate it altogether. The function we use to fit $G(\Delta j)$ reads\cite{Mussardo}
\begin{equation}
\label{G:fit}
G_{\rm fit}(\Delta j) = \alpha \,\Delta j^{-\eta} \exp(-\Delta j/\lambda)\,, 
\end{equation}
where $\alpha$, $\eta$ and $\lambda$ are fitting parameters. Figure~\ref{fig:correl} displays two-point correlations
functions $G(\Delta j)$ numerically evaluated
at distances up to 300 sites, and the fit according to Eq. \eqref{G:fit}, which provides in all the scenarios considered an impressive match to the numerical data. 

After sampling results for several points of the critical boundary, we obtain an average critical exponent of $\eta = 0.258 \pm 0.012$, in good agreement with the predicted $1/4$ value.

\begin{figure}
 \begin{center}
 \begin{overpic}[width = \columnwidth, unit=1pt]{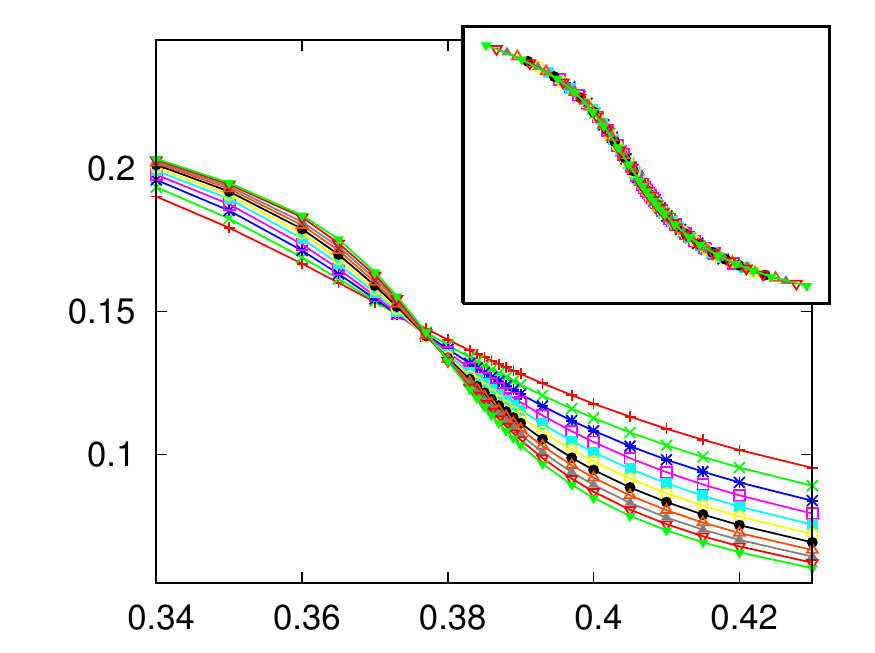}
  \put(20, 114){\large $\xi$}
  \put(130, -8){\large $\tilde{\omega}$}
  \put(180, 145){$f(x)$}
 \end{overpic}
 \end{center}
\caption{ \label{fig:finsiz} (color online)
Order parameter $\xi_L(g, \tilde{\omega})$, obtained via DMRG,
as a function of $\tilde{\omega}$, plotted for different system sizes $L = 100, 120 \ldots 300$, at $g=0.12$.
Inset: rescaled data
$\xi_L(\tilde{\omega}_c + \Delta\tilde{\omega}_c L^{\gamma_2}) L^{\gamma_1}$,
with $\gamma_1 = 0.127$ and $\gamma_2 = 1.04$,
characterizing $f(x)$ (see text).
}
\end{figure}

\subsection{Finite-size scaling}

By means of finite-size scaling \cite{FishBarb} it is possible to acquire two critical exponents:
the spontaneous magnetization exponent $\beta$ and the correlation length divergence exponent $\nu$.
In accordance to renormalization group analysis, the order parameter $\xi_L(g, \tilde{\omega})$ in an OBC setup
should obey a precise scaling with the system size $L$ and the parametric distance from the critical point
$(\tilde{\omega} - \tilde{\omega}_c)$. Specifically
\begin{equation} \label{eq:fishbarb}
 \xi_L (\tilde{\omega}) \simeq L^{-\beta / \nu} \; f \left( (\tilde{\omega} - \tilde{\omega}_c) \; L^{1 / \nu} \right),
\end{equation}
where $f$ is a non-universal function, depending on the microscopic details of the model.
To exploit this picture, first we tune $\gamma_1 = \beta / \nu$ until we observe a crossing of all the curves
$\xi_L(g, \tilde{\omega}) \; L^{\gamma_1}$
as functions of $\tilde{\omega}$ in a single point, which locates the critical frequency $\tilde{\omega}_c(g)$.
Then, we find the suitable value $\gamma_2 = - 1/\nu$ which makes all the curves
$\xi_L(g, \tilde{\omega}_c + (\tilde{\omega}-\tilde{\omega}_c)L^{\gamma_2}) \, L^{\gamma_1}$ collapse onto one another.
Fig.~\ref{fig:finsiz} shows an example of this procedure applied to our problem:
after plotting the order parameter $\xi_L$ as a function of the trap frequency $\tilde{\omega}$ for a dozen different
systems lengths $100 \leq L \leq 300$, we rescale the coordinates of the data according to this procedure,
and obtain the plot shown in the inset, where all the data collapsed onto a single curve,
which is the $f(x)$ from Eq.~\eqref{eq:fishbarb}.
Critical exponents obtained by employing this strategy
read $\beta = 0.126 \pm 0.011$ and $\nu = 1.03 \pm 0.05$, to be compared with theoretical
predictions of $1/8$ and $1$ respectively.

\subsection{Central charge $c$}

\begin{figure}
 \begin{center}
 \begin{overpic}[width = \columnwidth, unit=1pt]{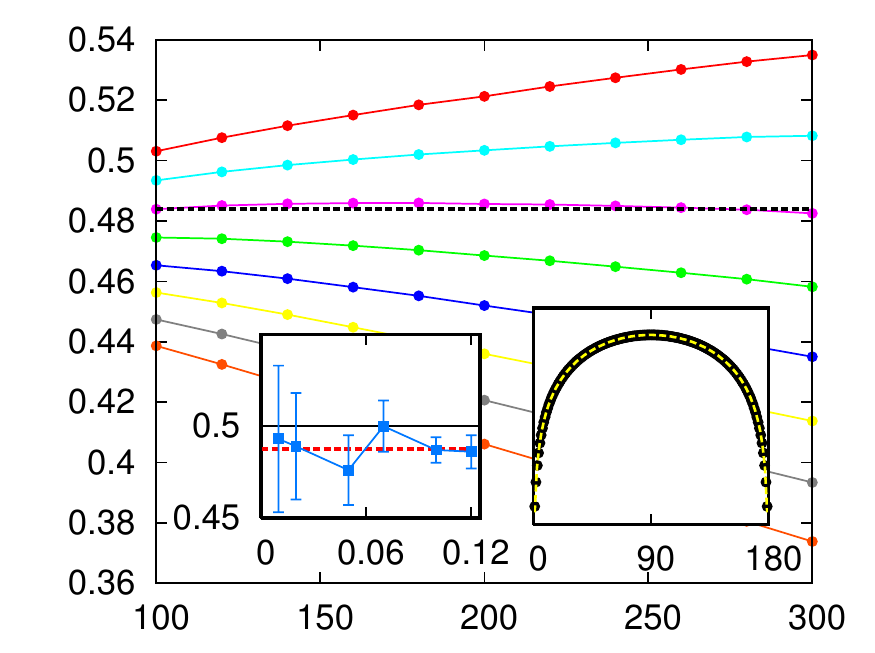}
  \put(4, 115){\large $c$}
  \put(154, -4){\large $L$}
  \put(198, 24){$\ell$}
  \put(85, 27){$g$}
  \put(60, 48){$c$}
  \put(170, 61){\large $\eS(\rho_\ell)$}
 \end{overpic}
 \end{center}
\caption{\label{fig:central}
Right inset: von Neumann entropy $\eS(\rho_{\ell})$ (black dots) of the reduced density matrix $\rho_{\ell}$
calculated applying Eq.~\eqref{eq:vonneudef} to the ground state simulation with DMRG,
as a function of the block size $\ell$.
The data shows remarkable agreement with the fit using $\eS_{\text{th}}(\rho_{\ell})$ of
Eq.~\eqref{eq:cardy} via $c$ and $c'$ (dashed yellow line). Main: $c$ values as a function of the system size
$L = 100,120 \ldots 300$ for different trap frequencies $\tilde{\omega}^2 = 0.383, 0.384 \ldots 0.390$
(top to bottom) at $g = 0.12$.
The magenta data set detects the critical point
at $\tilde{\omega}_c^2 = 0.385$, and the corresponding estimated $c$ value (dashed black line)
is $c = 0.486$.
Left inset: estimated central charge $c$ values at various points in the critical boundary,
as a function of the parameter $g$. The dashed line corresponds to the average of $c = 0.487$ while
the solid line is the theoretical prediction $c_{\text{th}}=1/2$.
}
\end{figure}
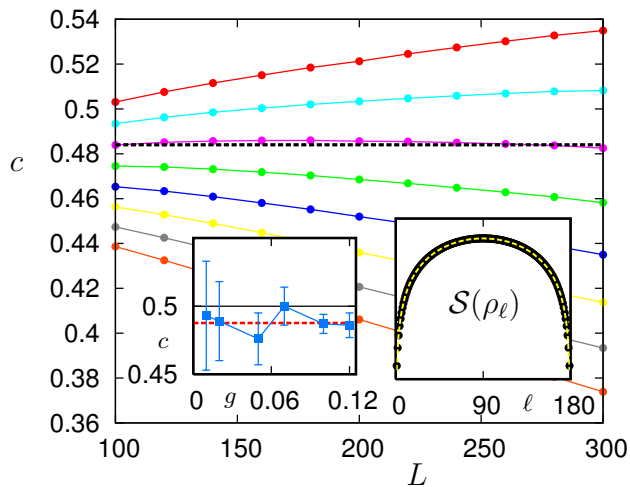

The central charge critical exponent $c$
is related to the scaling of quantum entanglement, under a system bipartition
into two blocks, as a function of the shape and size of the blocks themselves
\cite{Cencharge, VonNeu}.
Indeed, for one-dimensional systems, a logarithmic violation of the area law of entanglement
is a discriminating signature for criticality and a gapless excitation spectrum\cite{Cardy}.
In this framework, the central charge $c$ is the prefactor of the logarithm itself:
at the thermodynamical limit it reads
$\eS(\rho_\ell) \simeq \frac{c}{3} \log \ell + c'$, where
$c'$ a non-universal constant\cite{dechiara06}, and
\begin{equation} \label{eq:vonneudef}
 \eS(\rho_\ell) \equiv - \trace [\rho_\ell \ln \rho_\ell]
\end{equation}
is the Von Neumann entropy of the reduced density matrix
$\rho_\ell$ of $\ell$ adjacent sites.
For a semi-infinite system instead, where the $\ell$ sites contain the single boundary, it is
$\eS(\rho_\ell) \simeq \frac{c}{6} \log \ell + c'$. Our scenario is a finite-size OBC system, for which the relation holds \cite{Calacardy}
\begin{equation} \label{eq:cardy}
  \eS_{\text{th}}(\rho_\ell) = \frac{c}{6} \log \left(L \cdot \sin{\frac{\pi \ell}{L}} \right) + c'\;,
\end{equation}
where $\ell$ is the site at which we are considering a left-right system bipartition, and $\rho_\ell$
is the reduced density matrix of the left (or right) block. Due to the intrinsic nature of
DMRG it is straightforward to evaluate it for every partition $\ell$,
since the Schmidt coefficients are automatically provided from the algorithm.
We fit numerical data using Eq.~\eqref{eq:cardy} via $c$ and $c'$ for various system lengths $L$,
as shown in Fig.~\ref{fig:central}, right inset. At the critical point Eq.~\eqref{eq:cardy} should reproduce
the correct scaling of the entanglement, and thus the fitted $c$ values should be constant as a function of $L$.
This is actually the case, as shown in Fig.~\ref{fig:central}, main plot:
the magenta line corresponds to the critical point and is roughly constant with $L$.
This procedure provides an additional method for locating the critical point,
albeit less precise than the previously discussed ones, and quantifies the critical central charge at the same time.
After averaging over several points of the critical boundary (Fig.~\ref{fig:central}, left inset),
we acquire an estimate of $c = 0.487 \pm 0.015$, in good accordance with the theoretical value of $1/2$.

\section{Conclusions}\label{sec:conclu}

In this work we numerically studied the quantum linear-zigzag transition in quasi one-dimensional Wigner crystals by means of extensive numerical simulations based on Density Matrix Renormalization Group.

We first introduced the theoretical framework which lets us describe the phenomenon as a quantum lattice model, as well as the approximations we employed to make it amenable to simulation with DMRG.
According to this picture, we provided an analytical mapping from the original long-range theory to an effective short-range model.
We argued that such a mapping is possible as long as the excitations energies in the many-body state
are small compared to the phononic bandwidth $\eM_1$.
Applying this mapping leads to a critical speed-up of the numerical calculation, as we could employ traditional
nearest-neighbor model DMRG techniques to an otherwise difficult problem.

The phase diagram of the phase transition in the reduced external parameter space was determined. We detected the universality class of the criticality by extracting several critical exponents with high precision. These results ultimately show that the linear-zigzag transition is of the same universality class as the quantum Ising model in transverse field.  

Our model can be applied to strongly-interacting systems, such as ions in traps, dipolar gases, and Rydberg atoms chains: In fact, as long as the interaction is repulsive and described by a power law, the exponent $\alpha$ determining the strength of the interparticle potential just enters the coefficients of the transverse Hamiltonian, but does not change its form (as long as $\alpha\ge 1$). On the other hand, the study we performed is based on the assumption that transverse and longitudinal vibrations are decoupled. This assumption is correct for ion Coulomb systems, but has to be checked for dipolar and Rydberg systems, where the effect of quantum fluctuations may modify the nature of the transition \cite{Meyer-Rosch}.

\emph{Acknowledgments - } We acknowledge support from EU through PICC, SIQS, and from the German Research Foundation (Heisenberg programme, SFB/TRR21), the BW-grid for computational resources. We warmly thank G. De Chiara, S. Fishman, A. Muramatsu, M.B. Plenio, A. Retzker, A. Rosch, E. Shimshoni, D. Fioretto, and M. Burrello for discussions, and D. Rossini for contributing in the development of the numerical suite. 

\appendix

\begin{figure}
 \begin{center}
 \begin{overpic}[width = \columnwidth, unit=1pt]{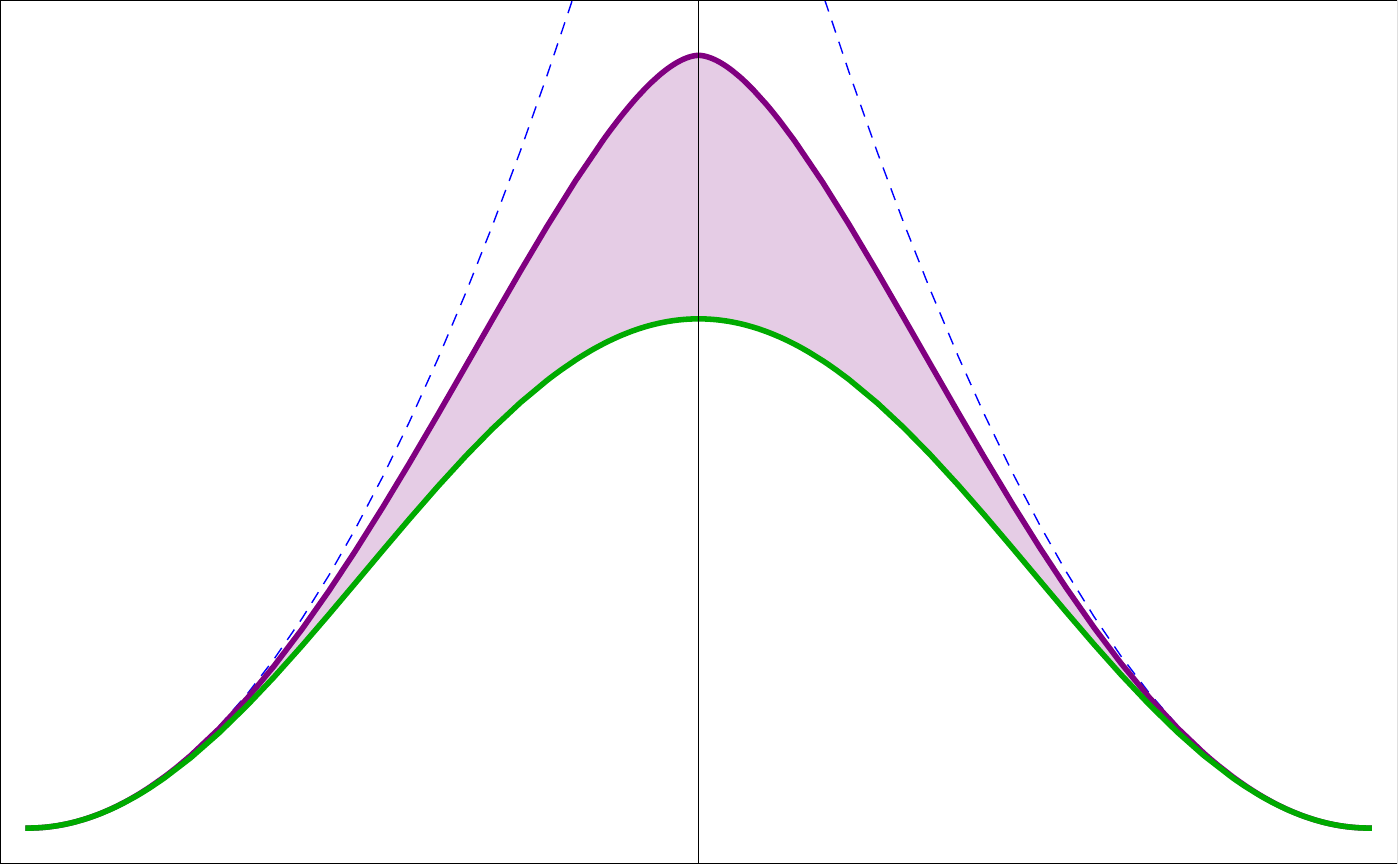}
  \put(241, -8){$\pi$}  
  \put(0, -8){-$\pi$}  
 \end{overpic}
 \end{center}
\caption{ \label{fig:bandstruct1}
(Color online) Original long-range single-phonon (\ie $q=1$) band profile $\Xi[k_1]$
of Eq.~\eqref{eq:scatter}, for $\alpha = 1$ (purple curve).
The parabola (dashed line) is the second-order expansion around the soft mode $k = \pi$, the green curve is the dispersion relation for the corresponding model with nearest-neighbor interaction, and matches the original dispersion band up to $\delta k^2$ around $\pi$.}
\end{figure}

\section{Short-range mapping formal derivation}\label{sec:append}

Here we are going to sketch the algebraic technique that we adopt to recast the long-range interacting model
into a nearest-neighbor interacting theory, provided that we want to faithfully reproduce the structural transition phenomenon.
In particular, the following framework will treat every term in the displacement expansion of Eq.~\eqref{eq:truexpand}
independently, and renormalize it into a finite-range object.
For this reason we recast Hamiltonian $\tilde{H}$, Eq \eqref{eq:truexpand}, as $\tilde{H} = H_0 + \frac{1}{2} \sum_q W_q$, where $H_0 = \frac{1}{2} \sum_j^L [\tilde{p}^2_j + \tilde{\omega}^2 \tilde{y}_j^2]$ is the local part and 
\begin{equation} \label{eq:monoterm}
 W_q = (-1)^q \; b_q(\alpha)
 \sum_{j = 1}^{L} \sum_{\ell \neq 0} \frac{1}{|\ell|^{\alpha+2q}}
 (\tilde{y}_{j+\ell} - \tilde{y}_{j})^{2q}
\end{equation}
is the $q$-th order of the Taylor expansion of the Coulomb interaction. In order to reveal the role played by the soft mode within $W_q$, we move to Fourier space. Therefore we temporarily assume periodic boundary conditions and move to the
discrete complex Fourier basis
\begin{equation} \label{eq:fourier}
 \tilde{y}_j = \frac{1}{\sqrt{L}} \sum_{k} Y_k e^{\imath j k}\,,
 \qquad
 Y_k = \frac{1}{\sqrt{L}} \sum_{j = 1}^L \tilde{y}_j e^{-\imath j k}\,,
\end{equation}
\begin{figure*}
 \begin{center}
 \begin{overpic}[width = \textwidth, unit=1pt]{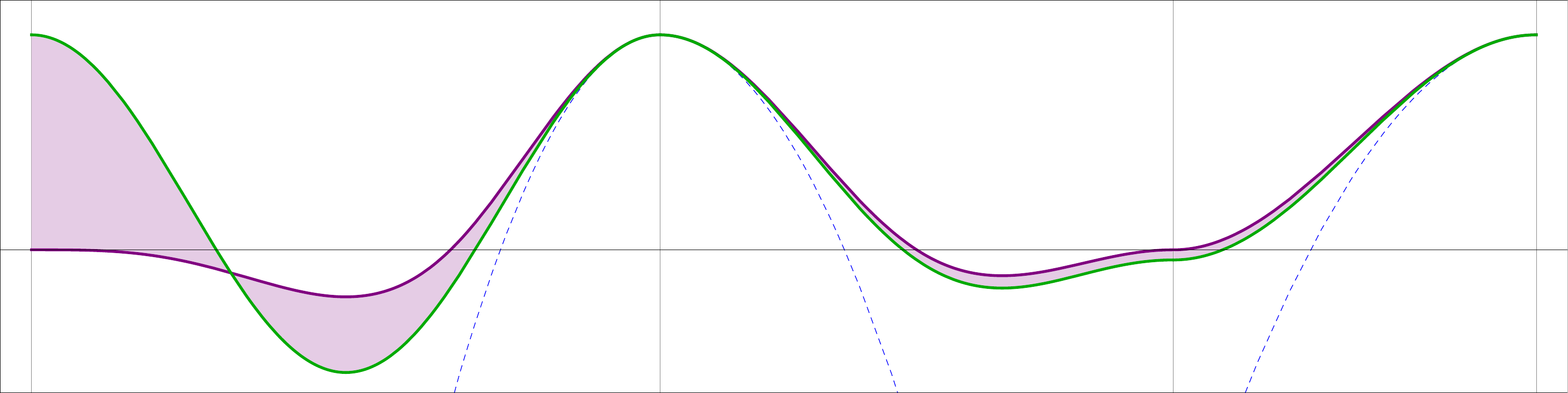}
  \put(8, -10){$\Gamma$}
  \put(212, -10){$\vec{\pi}$}
  \put(371, -10){$X \,|\, M$}
  \put(497, -10){$\vec{\pi}$}
 \end{overpic}
 \end{center}
\caption{ \label{fig:bandstruct2}
(color online) $q = 2$ scattering function $\Xi[k_1, k_2, k_3]$ of Eq.~\eqref{eq:scatter}, along some high-symmetry lines of the cubic Brillouin zone, of the original long-range model (purple line), encompassing
$\Gamma = (0,0,0)$ (the center of the cube), $X = (0,0,\pi)$ (the face center), and $M =(0,\pi,\pi)$ (the edge center).
The dashed line is the second-order expansion around
the soft mode point $\vec{\pi}$; the green line is the effective short-range model.
}
\end{figure*}
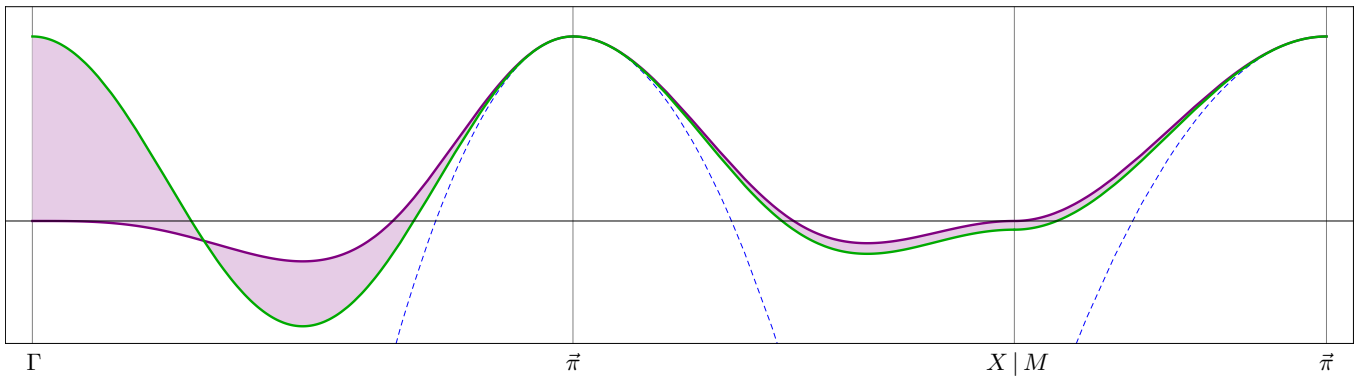
where $L$ is the total chain length, and the allowed $k$ values belong to the Brillouin zone grid, i.e., they are of the form $k = \pm \pi n / L$, with $|n|\leq N/2$ and integer
(again in tight-binding lattice units $a = 1$).
Plugging this substitution inside
\eqref{eq:monoterm} yields
\begin{equation} \label{eq:presca}
 W_q = \frac{2^{2q+1}}{L^{q-1}} \,b_q\!\!\!  \sum_{k_1 \ldots k_{2q-1}} \!\!
 Y_{k_1} \ldots Y_{k_{2Q-1}} Y_{-\!\sum\! k} \;\Xi [\vec{k} ],
\end{equation}
where each one of the $k_a$ values is chosen in the Brillouin grid.
Let us express the array of $\{k_a\}$ values as a $2q-1$ dimensioned vector $\vec{k}$,
and the scattering function $\Xi[\vec{k}]$ reads
\begin{equation} \label{eq:scatter}
 \Xi [\vec{k} ] = - \sum_{\ell > 0} \frac{1}{\ell^{\alpha+2Q}} \;
 \sin\left(\frac{\sum\! k}{2} \ell \right) \prod_{t = 1}^{2Q - 1} \sin\left(\frac{k_t \ell}{2}\right).
\end{equation}
In Fig.~\ref{fig:bandstruct1} and Fig.~\ref{fig:bandstruct2} we show two examples of scattering functions,
in the ion chain scenario $\alpha = 1$:
respectively we show $\Xi [k]$ for $q = 1$ in Fig.~\ref{fig:bandstruct1}, purple curve,
and $\Xi [\vec{k} ]$ for $q = 2$ along some high-symmetry lines of the Brillouin zone in Fig.~\ref{fig:bandstruct2}, purple curve.
So far the treatment is exact, now we make some approximations. Since we are only interested in the mechanics
related to the transition, which is primarily driven by the soft mode $k_0 = \pi$, we wish to keep only two orders
of Taylor expansion of $\Xi [\vec{k} ]$ around $\vec{\pi} = (\pi, \ldots, \pi)$. We will show in this section that it is possible to build a nearest-neighbor interaction term which matches the scattering function up to the second order.

First of all, let us calculate the expansion coefficients. The zeroth order reads
\begin{equation}
 \Xi [\vec{\pi} ] = (-1)^{q} \;\frac{2^{\alpha+2q}-1}{2^{\alpha+2q}} \;\zeta(\alpha + 2q),
\end{equation}
with $\zeta(\cdot)$ being the Riemann zeta function. The second order must be treated carefully: here we will restrict our study to terms with $q \geq 2$.
The first term $q = 1$ requires a slightly different treatment, which is reported in Ref. \cite{Morigi:G.5}. Let us now write the Hessian matrix of the scattering function $\Xi$ in $\vec{\pi}$,
by respectively writing the non-diagonal elements
\begin{equation}
 \left. \frac{\partial^2 \Xi [\vec{k} ]}{\partial k_a \partial k_{b \neq a}} \right|_{\vec{\pi}}
 = (-1)^{q+1} \;\frac{2^{\alpha+2q-2}-1}{2^{\alpha+2q}} \;\zeta(\alpha + 2q - 2),
\end{equation}
and the diagonal ones, whose value is exactly twice as much
\begin{equation}
 \left. \frac{\partial^2 \Xi [\vec{k} ]}{\partial k_a^2} \right|_{\vec{\pi}} = 2 \cdot
 \left. \frac{\partial^2 \Xi [\vec{k} ]}{\partial k_a \partial k_{b \neq a}} \right|_{\vec{\pi}}
\end{equation}
Now since the $Y_k$ operators commute,
by a simple relabeling of $k_a$ indices in expression \eqref{eq:presca},
it is possible to reshape the Hessian: as long as the diagonal and off-diagonal element are not mixed,
the sum in \eqref{eq:presca} is left unchanged.
More precisely, we reshape the matrix $T_{a b} =\partial_{k_a} \partial_{k_b} \Xi[\vec{\pi}]$ as follows. Starting from
\begin{equation}
 T =\left( \begin{array}{cccc}
 2 & 1 & \ldots & 1 \\
 1 & 2 & \ldots & 1 \\
 \vdots & \vdots & \ddots & \vdots \\
 1 & 1 & \ldots & 2 \\
 \end{array} \right)
 \quad \mbox{dimension } 2q-1
\end{equation}
we end up with
\begin{equation}
 T' = 2 \frac{2q - 1}{q}
 \left( \begin{array}{ccc}
 1 & \ldots & 1 \\
 \vdots & \ddots & \vdots \\
 1 &  \ldots & 1 \\
 \end{array} \right)
 \quad \mbox{dimension } q.
\end{equation}
Therefore, including the $1/2!$ factor from the second order Taylor expansion in $k$,
we obtain
\begin{multline} \label{eq:almosthere}
 W_q \simeq \frac{1}{2 L^{q-1}} \!\!\!  \sum_{k_1 \ldots k_{2q-1}} \!\!
 Y_{k_1} \ldots Y_{k_{2q-1}} Y_{-\!\sum\! k}
 \times \\ \times
 (-1)^q
 \left(
 \eM_q - \eN_q \cdot (k_1 + \ldots + k_{q} - q \pi)^2 \right)
\end{multline}
with $\eM_q$ and $\eN_q$ coefficients given by Eqs.~\eqref{eq:onsitefield} and \eqref{eq:offsitefield}.
As last step, we substitute the expression $(k_1 + \ldots + k_{q} - q \pi)^2$ inside Eq.~\eqref{eq:almosthere}
with a cosine function which, again, matches two Taylor orders in $k$ around $\vec{\pi}$,
so that we are not increasing the error order, \ie we are not adding further approximation. Precisely
\begin{equation}
 \left[ K - q\pi \right]^2 \to 2\left(1 - (-1)^q \cos K\right)
\end{equation}
with $K = k_1 + \ldots + k_q$. If we now transform back into real space we end up with a nearest-neighbor interaction term,
which reads
\begin{equation} \label{eq:funny}
 W_q \simeq \frac{1}{2} \sum_{j = 1}^{L}
 \left[ \eM_q(\alpha) \,\tilde{y}_j^{2q} - \mathcal{N}_q(\alpha) \left( \tilde{y}_j^{q} - (- \tilde{y}_{j+1} )^{q} \right)^2
 \right]
  \;.
\end{equation}
This corresponds exactly to Eq.~\eqref{eq:serious} and thus concludes the treatment.
The green lines in Fig.~\ref{fig:bandstruct1} and Fig.~\ref{fig:bandstruct2} show how the
scattering functions $\Xi[\vec{k}]$ are modified according to this approximation: the match
around the soft mode point $\vec{k} = \vec{\pi}$ is evident.

We remark that this procedure extends the analisys
introduced in Refs. \cite{DeChiara:2008,Shimshoni:2011b}, where the first two Taylor expansion terms $W_1$ and $W_2$
were considered in the mapping from long to short range, for
arbitrary power-law scaling $\alpha \geq 1$ of the original two-body interaction.
Equations \eqref{eq:onsitefield}, \eqref{eq:offsitefield} and \eqref{eq:funny}
generalize this method to every $q \to q$ phononic scattering channel.


\end{document}